\documentclass[aps,prb,amsmath,amssymb,amsfont,showpacs,superscriptaddress,reprint]{revtex4-1}

\usepackage{hyperref}
\usepackage{ifpdf}

\usepackage{times}
\usepackage{graphicx}
\usepackage[usenames,dvipsnames]{xcolor}
\usepackage{array}
\usepackage{tabularx}

\begin{document}

\title{Incidence of multilayers in chemically exfoliated graphene}

\author{P.~Szirmai}
\thanks{These two authors contributed equally.}
\affiliation{Faculty of Physics, University of Vienna, Strudlhofgasse 4., Vienna, A-1090, Austria}
\affiliation{Laboratory of Physics of Complex Matter, \'{E}cole Polytechnique F\'{e}d\'{e}rale de Lausanne, Lausanne CH-1015, Switzerland}
\affiliation{Department of Physics, Budapest University of Technology and Economics and MTA-BME Lend\"{u}let Spintronics Research Group (PROSPIN), PO Box 91, H-1521 Budapest, Hungary}

\author{B.~G.~M\'{a}rkus}
\thanks{These two authors contributed equally.}
\affiliation{Faculty of Physics, University of Vienna, Strudlhofgasse 4., Vienna, A-1090, Austria}
\affiliation{Department of Physics, Budapest University of Technology and Economics and MTA-BME Lend\"{u}let Spintronics Research Group (PROSPIN), PO Box 91, H-1521 Budapest, Hungary}

\author{J.~C.~Chac\'{o}n-Torres}
\affiliation{Yachay Tech.~University, School of Physical Sciences and Nanotechnology, 100119 Urcuqu\'{i}, Ecuador}
\affiliation{Institut f\"{u}r Experimental Physik, Freie Universit\"{a}t Berlin, Arnimallee 14, 14195 Berlin, Germany}

\author{P.~Eckerlein}
\affiliation{Department of Chemistry and Pharmacy and Joint Institute of Advanced Materials and Processes (ZMP), Friedrich-Alexander University of Erlangen-N\"{u}rnberg, Nikolaus-Fiebiger-Str.~10, 91058 Erlangen, Germany}

\author{K.~Edelthalhammer}
\affiliation{Department of Chemistry and Pharmacy and Joint Institute of Advanced Materials and Processes (ZMP), Friedrich-Alexander University of Erlangen-N\"{u}rnberg, Nikolaus-Fiebiger-Str.~10, 91058 Erlangen, Germany}

\author{J.~M.~Englert}
\affiliation{Department of Chemistry and Pharmacy and Joint Institute of Advanced Materials and Processes (ZMP), Friedrich-Alexander University of Erlangen-N\"{u}rnberg, Nikolaus-Fiebiger-Str.~10, 91058 Erlangen, Germany}

\author{U.~Mundloch}
\affiliation{Department of Chemistry and Pharmacy and Joint Institute of Advanced Materials and Processes (ZMP), Friedrich-Alexander University of Erlangen-N\"{u}rnberg, Nikolaus-Fiebiger-Str.~10, 91058 Erlangen, Germany}

\author{A.~Hirsch}
\affiliation{Department of Chemistry and Pharmacy and Joint Institute of Advanced Materials and Processes (ZMP), Friedrich-Alexander University of Erlangen-N\"{u}rnberg, Nikolaus-Fiebiger-Str.~10, 91058 Erlangen, Germany}

\author{F.~Hauke}
\affiliation{Department of Chemistry and Pharmacy and Joint Institute of Advanced Materials and Processes (ZMP), Friedrich-Alexander University of Erlangen-N\"{u}rnberg, Nikolaus-Fiebiger-Str.~10, 91058 Erlangen, Germany}

\author{B.~N\'{a}fr\'{a}di}
\affiliation{Laboratory of Physics of Complex Matter, \'{E}cole Polytechnique F\'{e}d\'{e}rale de Lausanne, Lausanne CH-1015, Switzerland}

\author{L.~Forr\'{o}}
\affiliation{Laboratory of Physics of Complex Matter, \'{E}cole Polytechnique F\'{e}d\'{e}rale de Lausanne, Lausanne CH-1015, Switzerland}

\author{C.~Kramberger}
\affiliation{Faculty of Physics, University of Vienna, Strudlhofgasse 4., Vienna, A-1090, Austria}

\author{T.~Pichler}
\affiliation{Faculty of Physics, University of Vienna, Strudlhofgasse 4., Vienna, A-1090, Austria}

\author{F.~Simon}
\affiliation{Department of Physics, Budapest University of Technology and Economics and MTA-BME Lend\"{u}let Spintronics Research Group (PROSPIN), PO Box 91, H-1521 Budapest, Hungary}

\begin{abstract}
An efficient route to synthesize macroscopic amounts of graphene is highly desired and a bulk characterization of such samples, in terms of the number of layers, is equally important. We present a Raman spectroscopy-based method to determine the distribution of the number of graphene layers in chemically exfoliated graphene. We utilize a controlled vapor-phase potassium intercalation technique and identify a lightly doped stage, where the Raman modes of undoped and doped few-layer graphene flakes coexist. The spectra can be unambiguously distinguished from alkali doped graphite, and a modeling with the distribution of the layers yields an upper limit of flake thickness of five layers with a significant single-layer graphene content. Complementary statistical AFM measurements on individual few-layer graphene flakes find a consistent distribution of the layer numbers.
\end{abstract}

\maketitle

\section{Introduction}

Graphene, the latest discovered carbon allotrope~\cite{NovoselovSCI2004,GeimRMP2009}, holds promise for a wide range of potential applications from medical devices to sensors~\cite{GeimSCI2009,BonaccorsoNatPh2010,Mikhailov2011,Avouris2012}. Apart from individual graphene flakes, bulk graphene is exploited in numerous systems~\cite{FerrariNanoscale2015,Raccichini2015}. Notably, bulk single-layer and few-layer graphene (SLG and FLG) are proposed as applicable in efficient Li-ion batteries, components of photovoltaic cells, and are viable candidates for spintronics applications~\cite{TombrosNAT2007,YangPRL2011,FerrariNanoscale2015}. Thus, scalable methods are required both for high-yield production and characterization techniques. One of the major challenges, i.e., to establish mass-production techniques leading to high-quality SLG and FLG was overcome in recent years~\cite{VallesJACS2008,EiglerAngew2014}. Amongst the numerous synthesis (mostly top-down) means towards mass graphene production, wet chemical exfoliation methods prevail in terms of material quality and synthesis facility~\cite{HernandezNatNano2008,LotyaJACS2009,ParedesJMatChem2011,ColemanAChemRes2013}. Nevertheless, measurement methods suitable for large sample quantities are still lacking.

Raman spectroscopy evolved to be an essential probe for studies of carbon structures especially of nanocarbon~\cite{FerrariPRL2006, Ni2008a,Dresselhaus2010}. It was demonstrated to be an ideal characterization tool not only in laboratory but also at the mass-production level. In particular, investigations of single-layer graphene are facilitated by the lack of a band-gap in the band structure leading to resonant processes at all wavelengths~\cite{FerrariNatNano2013}. Details of the Raman spectra could not only reveal the number of the layer of single graphene flakes or identify the edge of the flakes~\cite{FerrariSSC2007,GrafSSC2007,GrafNanoLett2007,KohACSNano2011}, it also enables detailed studies of electrostatic gating, strain, or chemical modifications of graphene~\cite{Casiraghi2007a,Gupta2008,Das2009, Huang2009,Cong2010,Yoon2011,Cancado2011,Ni2008,Englert2011,Crowther2012,Mafra2012,Vecera2017}. 

Historically, Raman, phase contrast spectroscopy and AFM studies focused on the characterization of separated graphene flakes~\cite{FerrariPRL2006,Casiraghi2007,Malard2009}. Bulk analysis of the number of the layers of unmodified bulk graphene is not feasible by optical means. Raman spectra of SLG and FLG single flakes differ significantly in their 2D mode (the overtone of the defect induced D mode), which provides a way to identify the number of graphene layers, $N$~\cite{Park2009}. However, twisted multilayers might have 2D modes resembling to SLG~\cite{FerrariNatNano2013}. Furthermore, the methods reviewed in Ref.~\cite{FerrariNatNano2013} utilize the layer number dependence of the interlayer shear mode \cite{TanNatNano2012}, i.e. the so-called C peak and the interlayer breathing modes \cite{LuiNanoLett2012,LuiPRB2013,LuiNanoLett2014}, the ZO' modes. However, this cannot be applied for powder samples built up of a distribution of different thicknesses.

Here, we present a compelling analysis method based on Raman spectroscopy for analyzing the distribution of the number of graphene layers in a powder sample. We utilize controlled vapor-phase potassium intercalation to distinguish SLG and FLG content through following a stepwise intercalation process. This technique enables us to track the evolution of doping in lightly doped stages, where the Raman modes of the undoped and the doped FLG flakes coexist. The applicability of this intercalation method is demonstrated on a few-layer restacked graphene, poly-dispersed powder prepared by wet chemical exfoliation. The method reveals that the studied samples are composed of graphene flakes with maximum five layers. AFM statistical measurements on individual SLG and FLG flakes of the same material confirm the observed distribution.

\section{Methods}

Few-layer graphene samples were prepared from saturation potassium doped SGN18 spherical graphite powder (Future Carbon) using DMSO solvent for the wet chemical exfoliation as described elsewhere \cite{Englert2011,Vecera2014,VeceraNatCom2017} and as shown in Fig.~\ref{fig:scheme}\textit{a}. Chemical exfoliation was finalized using different mechanical routes: ultrasound sonication, shear mixing and mechanical stirring. In a previous study we showed that mechanical processing affects the material quality of final product~\cite{Markus2015}.

The as-prepared bulk samples were characterized by atomic force microscopy (AFM) topographic measurements and images to determine the flake size and the typical number of graphene layers in the bulk material. \textcolor{black}{For AFM statistical analysis, the samples were drop-casted on a 100$\times$100~$\mu$m surface of a Si/SiO$_2$ wafer. Unlike the samples used for Raman studies, only partial restacking may occur due to drop-casting and presence of the substrate in this configuration used for AFM studies.} These measurements were carried out on an NT-MDT AFM using a tungsten coated tip.

Potassium intercalation was carried out \emph{in situ} under a vacuum of $4 \times 10^{-8}$ mbar pressure. Prior to doping, the sample was resistively heated up to 500~$^{\circ}$C to evaporate the remaining solvent used before. The applied geometry is similar to the two-chamber vapor phase method used for graphite intercalated compounds (GICs)~\cite{Dresselhaus1981}. \textcolor{black}{Due to lack of substrate, significant restacking of individual graphene flakes occurs, leading to misaligned layers in the sample. The powder sample consists of sponge-like structures, as thermodynamic equilibrium is achieved to create a three-dimensional solid material. As a result of the described sample morphology, Raman measurements provide detailed information not only of single graphene flakes but a large ensemble of flakes that are restacked. As restacking creates only mechanical contact, electronic properties of individual flakes are preserved. Thus, the staging phenomenon necessary for our Raman-based technique occurs at the level of individual flakes.}

\textcolor{black}{Raman measurements were performed at $458$, $514$, and at $568$ nm laser excitation wavelengths. To facilitate our discussion, our manuscript focuses on the observations at $514$~nm wavelength, only the statistical analysis of the G-band positions in Figure~\ref{deltag_plot} is presented at all three wavelengths to confirm lack of dispersion. All other observations and our Raman-based method are independent of the wavelength, as well.}

A power of $0.5$~mW was used to avoid laser-induced deintercalation~\cite{Nemanich1977,ChaconTorres2012}. Potassium with a purity of $99.95\%$ (Sigma-Aldrich MKBL0124V) was evaporated to the sample in several steps in a controlled \emph{in situ} process. At each step, the potassium was evaporated for about $2$~minutes applying a decreasing temperature gradient, similarly to the well-known two-zone vapor phase method \cite{Dresselhaus1981}. Maximal doping was achieved in approximately $10$ steps in all cases. Near saturation doping a gradual color change was observed from black to brown and red, respectively. HOPG and single-crystal graphite intercalation compounds show a similar color change, however, no such significant change is apparent for graphite powder samples due to surface roughness \cite{Fabian2012}. This difference is seen as a proof of the smooth surface of the graphene layers in FLG \cite{SupMat}. 

Following each intercalation step, Raman spectra were recorded on a modified broadband LabRAM spectrometer (Horiba Jobin-Yvon Inc.). The built-in interference filter was replaced by a broadband beam splitter plate with $30\%$ reflection and $70\%$ transmission. The principles of the broadband operation are described elsewhere \cite{FabianRSI,FabianPSSB}. The spectrometer was operated with a $1800$ grooves$/$mm grating. A typical $0.5$~mW laser power was used with a built-in microscope (Olympus LMPlan $50\times/0.50$ inf$./0/$NN$26.5$).

\section{Results and discussion}

\begin{figure}[h!]
\includegraphics[width=\linewidth]{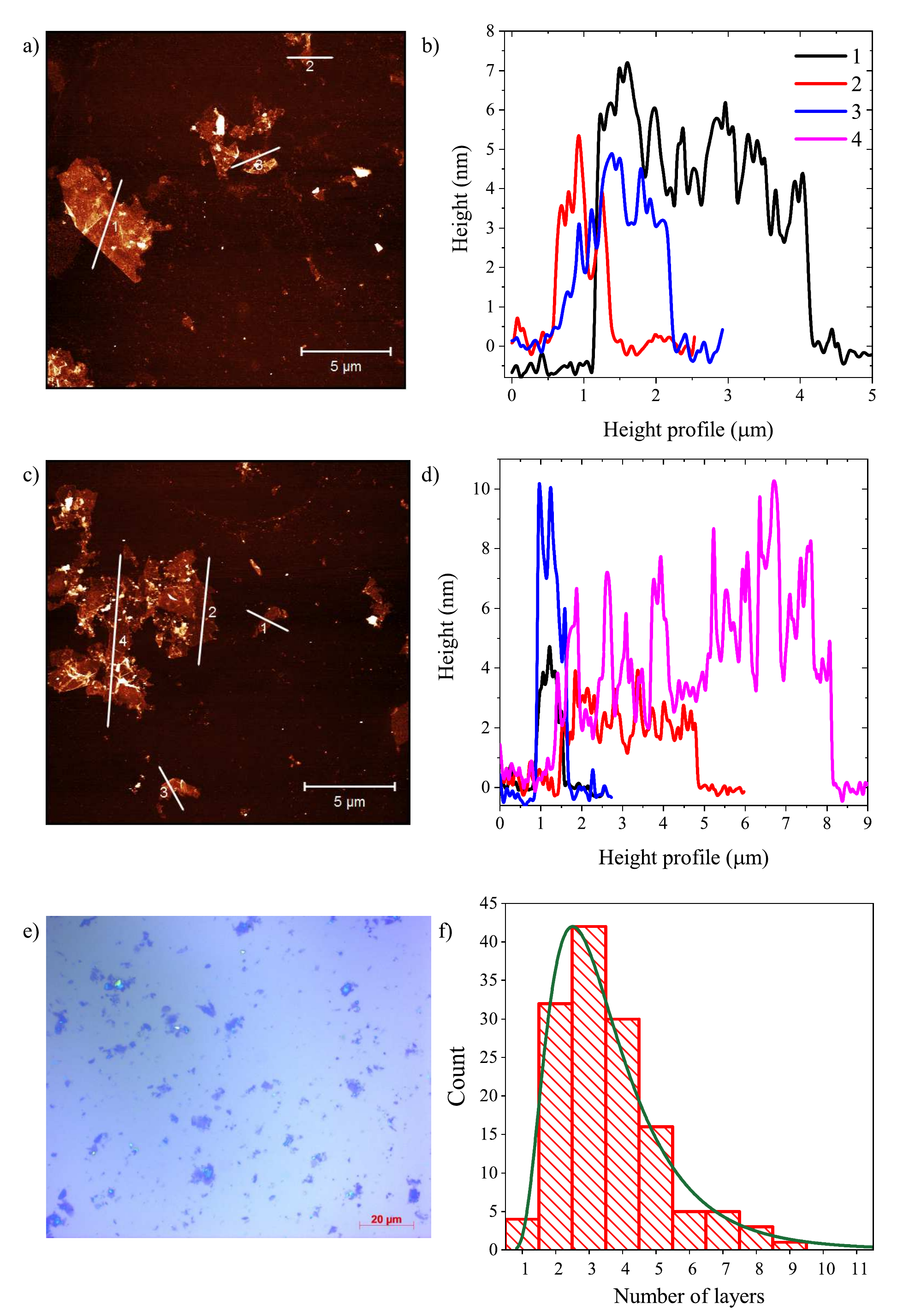}
\caption{\textcolor{black}{AFM experiments on chemically exfoliated few-layer graphene made with DMSO and ultrasound treatment. Multiple characteristic types of flakes can be identified: \emph{a)} and \emph{c)} show AFM studies containing few layer graphene sheets (up to $5$ layers). Note the diverse lateral size of the flakes that shows that these are partially restacked on the substrate. \emph{b), d)} Height profile of corresponding graphene flakes along the lines indicated in the left images. \emph{e)} Light microscope image depicting the distribution of flakes on a 100$\times$100~$\mu$m surface. \emph{f)} Distribution of flakes as a function of layer number in the AFM statistical analysis. Solid green line is a lognormal distribution fit to the height profiles revealing a mean of 3~layers for the thickness of flakes.}}
\label{fig:Fig1_AFM}
\end{figure}

\textcolor{black}{We performed a detailed AFM statistical analysis study on a large number of as-prepared ultrasound treated individual FLG flakes in a single batch chosen randomly. Fig.~\ref{fig:Fig1_AFM}\emph{e} shows a light microscope image on a 100$\times$100~$\mu$m surface, revealing a distribution of flakes on the surface. Representative AFM images of the graphene flakes are presented in Figure~\ref{fig:Fig1_AFM} along with cross-sectional cuts of the flakes. Figure~\ref{fig:Fig1_AFM}~\emph{a-d} point to presence of graphene flakes with up to five layers, with a sizeable fraction of mono-layer flakes.}

\textcolor{black}{Fig.~\ref{fig:Fig1_AFM}\emph{e} shows the distribution of flake thicknesses in our statistical analysis. This analysis highlights that ~90\% of the chemically exfoliated flakes are composed of maximum 5 graphene layers. A simple fit to a lognormal distribution points to a distribution of flakes centered at 3 layers (with a variance of 1.5 layers). Relevantly, a fraction of the flakes consists of single-layer graphene flakes in our sample.}

\textcolor{black}{Although AFM-based thickness measurement of individual flakes is a standard method for graphene characterization, it was suggested~\cite{VallesJACS2008,Nemes-Incze2008,Bonaccorso2012} that this approach may be misleading due to the improperly chosen measurement parameters, and complementary studies are required. In particular, partial restacking of the chemically exfoliated graphene flakes on the substrate may lead to bigger aggregates of multiple layers where the graphene sheets are misaligned. AFM, however, is unable to resolve this change of the flake morphology, and cannot identify the thickness of individual flakes.}

\textcolor{black}{This diversity of the flakes highlights the need for bulk characterization methods, such as Raman spectroscopy. Micro-Raman spectroscopy in our case has about a 5..10 $\mu$m lateral and vertical resolution, which is large enough for a representative surface average without the biasing effects of nano-imaging.}

\begin{figure}[h!]
\includegraphics*[width=\linewidth]{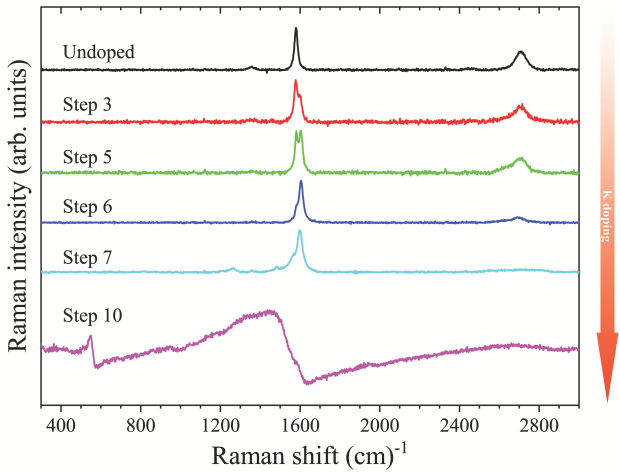}
\caption{Raman spectra of \emph{in-situ} potassium doped FLG starting from the undoped material (top) towards saturation doping (bottom). Saturation intercalation is reached after about $10$ intercalation steps, which are described in the text. Note that several steps are skipped in the figure that show little or no change. Upon doping, the D mode quickly disappears in accordance with previous literature data~\cite{Howard2011}. The 2D mode acquires some structure but also disappears after further intercalation steps. The G-band splits into G$_1$ and G$_2$, whose origin is discussed in the text. In the final, fully intercalated step, the G bands form a Fano-shaped band and a C$_{\textrm{z}}$-mode is observed at wavenumbers $\sim$560~cm$^{-1}$, similarly to Stage I graphite (KC$_8$).}
\label{fig:Fig2_flg_doping_all}
\end{figure}

Starting from undoped FLG, we performed controlled temperature-gradient driven potassium doping experiments. Saturation doping was achieved in approximately $10$ steps. We intentionally refer to "steps" in our experiments rather than "stages", as the latter is reserved for the well-known intercalation stages of bulk graphite \cite{Dresselhaus1981}. The corresponding Raman spectra (recorded at $514$ nm) are depicted in Figure~\ref{fig:Fig2_flg_doping_all}. Raman spectra of \textcolor{black}{the starting material} displays the usual D, G, and 2D bands and it reproduces the earlier report on similar samples \cite{Englert2011}. The 2D band of \textcolor{black}{the starting material} is best fitted with a single Lorentzian line, unlike the composed structure in graphite. Whereas the width of the Lorentzian hints that the material may be a mixture of flakes with different number of layers, the Raman response of \textcolor{black}{the starting material} is insufficient to determine the exact distribution of the thicknesses.

Upon light potassium doping, the Raman spectrum changes significantly: the weak D-band rapidly disappears and the G- and 2D-modes split. At higher doping levels, intensity of the double-resonant 2D peak components is suppressed, and both signals downshift. The highest doping level (Step~$10$ in Fig.~\ref{fig:Fig2_flg_doping_all}) leads to a radical change of the Raman spectrum. A Fano-shaped line~\cite{Fano1961,KuzmanyBook}, centered around $1486$~cm$^{-1}$, and a so-called C$_{\textrm{z}}$-like mode dominate the spectrum~\cite{ChaconTorres2012}. The Fano shape is a clear signature of significant charge transfer to the graphene sheets, which leads to a quantum interference of the zone-center phonons and the electronic transitions.

It is intriguing to compare this spectrum with the Raman spectrum of Stage I potassium intercalated graphite (KC$_8$), where similar Raman bands appear upon intercalation. A detailed analysis is given in the Supplementary Materials, and it indicates that the position, the width ($\Gamma_{\textrm{Fano}}$), and the coupling strength of the electronic continuum, measured by the asymmetry parameter, $q$, differ. Given that the C$_{\textrm{z}}$ mode arises from the C$-$K$-$C vibrational mode, it is reasonable that it is naturally present in Stage I KC$_8$. Here, it is only present in the multilayer flakes, and it is absent for the monolayer sheets~\cite{Howard2011}. This observation means that formation of a C$_{\textrm{z}}$-like mode is a clear indication of the presence of multilayer graphene flakes in the sample in a sizable amount.

The most surprising observation in Fig.~\ref{fig:Fig2_flg_doping_all} is the presence of a doublet G mode. In a sample, which contains SLG only, a homogeneous doping is expected to lead to a single G mode only. We can rule out the presence of an inhomogeneous doping~\cite{Howard2011} as we studied a large number of positions on the sample, several intercalation runs, and the same spectra were observed in all cases. It is however intriguing that the Raman spectrum of normal bulk graphite shows similar doublet structure under doping. In particular, the Raman spectrum of our Step 5 intercalated FLG may appear similar to a high stage (KC$_{72}$ or Stage VI) GIC. Nevertheless, the spectroscopic details are markedly different. 

\begin{figure}[h!]
\includegraphics*[width=\linewidth]{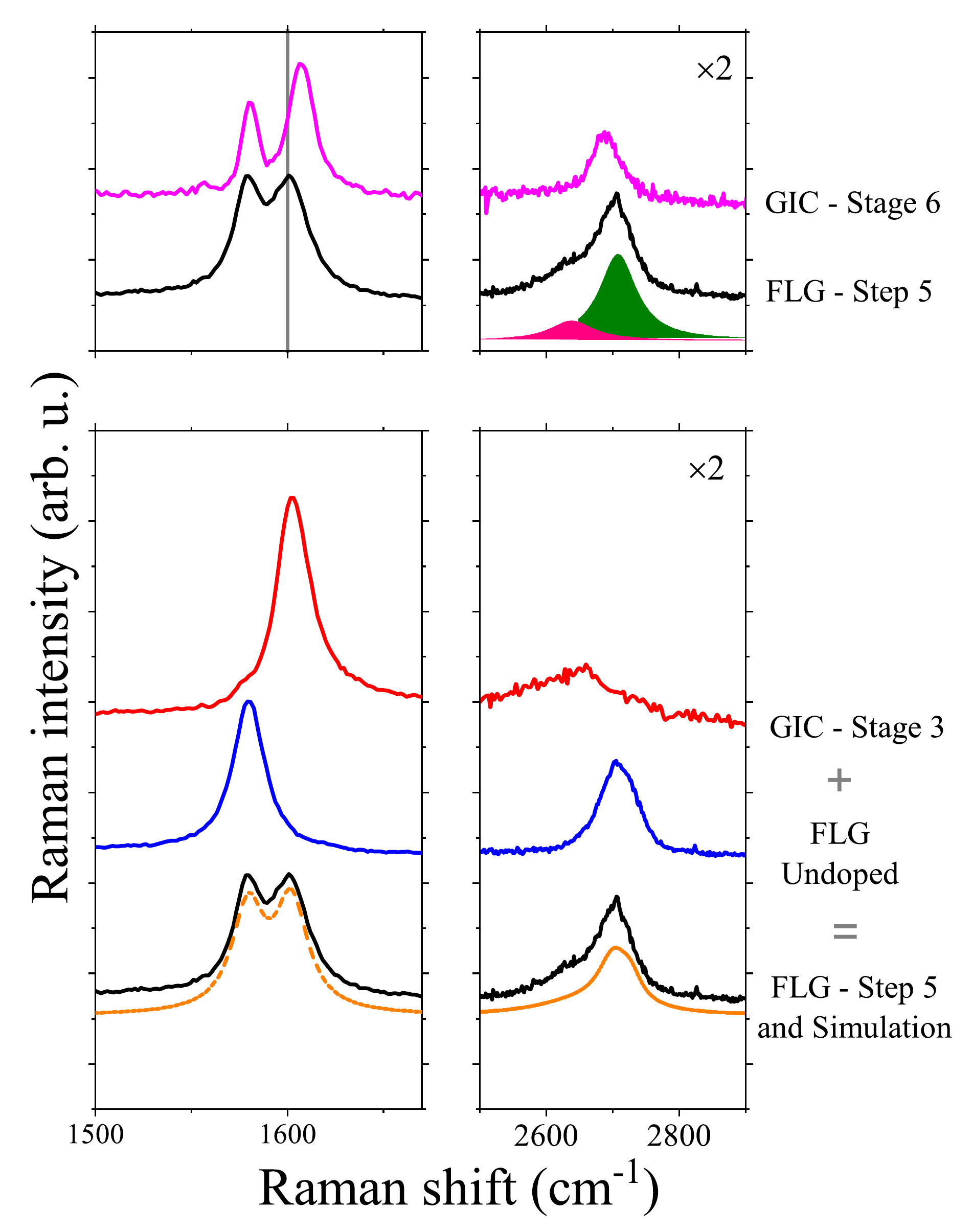}
\caption{Upper panel: Comparison of single-crystal graphite doped to Stage 6 and FLG doped to Step 5. Vertical line indicates the position of G$_2$ line in the doped FLG. A fit with two components (green and pink) simulates well the doped FLG signal. Lower panel: Simulation of the decomposition of the Raman spectrum of FLG doped to Step 5 as a mixture of a stage 3 GIC doped and the undoped FLG material. The bottommost spectrum is the simulated curve shown together with the Step 5 intercalated FLG (thus shown twice in the figure for clarity).}
\label{fig:simulation}
\end{figure}

In Fig.~\ref{fig:simulation}., we compare the Raman spectrum of our Step 5 intercalated FLG with the data on a graphite single crystal at a doping stage of 6 or KC$_{72}$. Although the doublet structure of the G mode appear similar, two details are different: i) the G mode with the larger Raman shift lies with about a $6(1)$ cm$^{-1}$ difference in the two types of materials, ii) the 2D mode is markedly different in the two kinds of materials: the FLG contains a 2D mode component with a smaller Raman shift, which is absent in graphite. Albeit these difference may appear to be subtle, these enable us to qualitatively differentiate between the two types of materials. The figure also shows that the Step 5 intercalated FLG can be resolved into a mixture of a Stage 3 GIC and the undoped material. \textcolor{black}{This fitting procedure (see Supplementary Section~\ref{fit_proc}) is capable of explaining both the position of the G mode and the composite structure of the 2D mode.} This indicates an interesting scenario for the Raman spectrum of the alkali intercalated FLG: it consists of a mixture of 1) entirely undoped pieces, whose Raman spectrum remains identical to that of the starting material, and 2) relatively highly doped phases (equivalent to Stage 3 GIC).

We emphasize that the origin of the doublet structure in GIC is related to the presence of charged and uncharged graphene layers; graphene layers in GIC, which are adjacent to an alkali layer are charged, whereas two which are further apart, remain neutral or uncharged~\cite{ChaconTorres2013}. In this respect, charges in GIC and and our stepwise doped FLG are both inhomogeneously distributed, however, the inhomogeneity is completely different. Intercalation in graphite proceeds from a homogeneous and crystalline graphite and the charging inhomogeneity is due to the intercalation itself: it occurs due to the thermodynamic preference for fully doped alkali layers which are inevitably separated by uncharged graphene layers. However in our FLG material, the inhomogeneity is present \emph{a priori} in the sample (in terms of the different layer numbers in the grains) and the inhomogeneous doping merely reflects this inhomogeneity as we show below.

\begin{figure}[h!]
\includegraphics*[width=\linewidth]{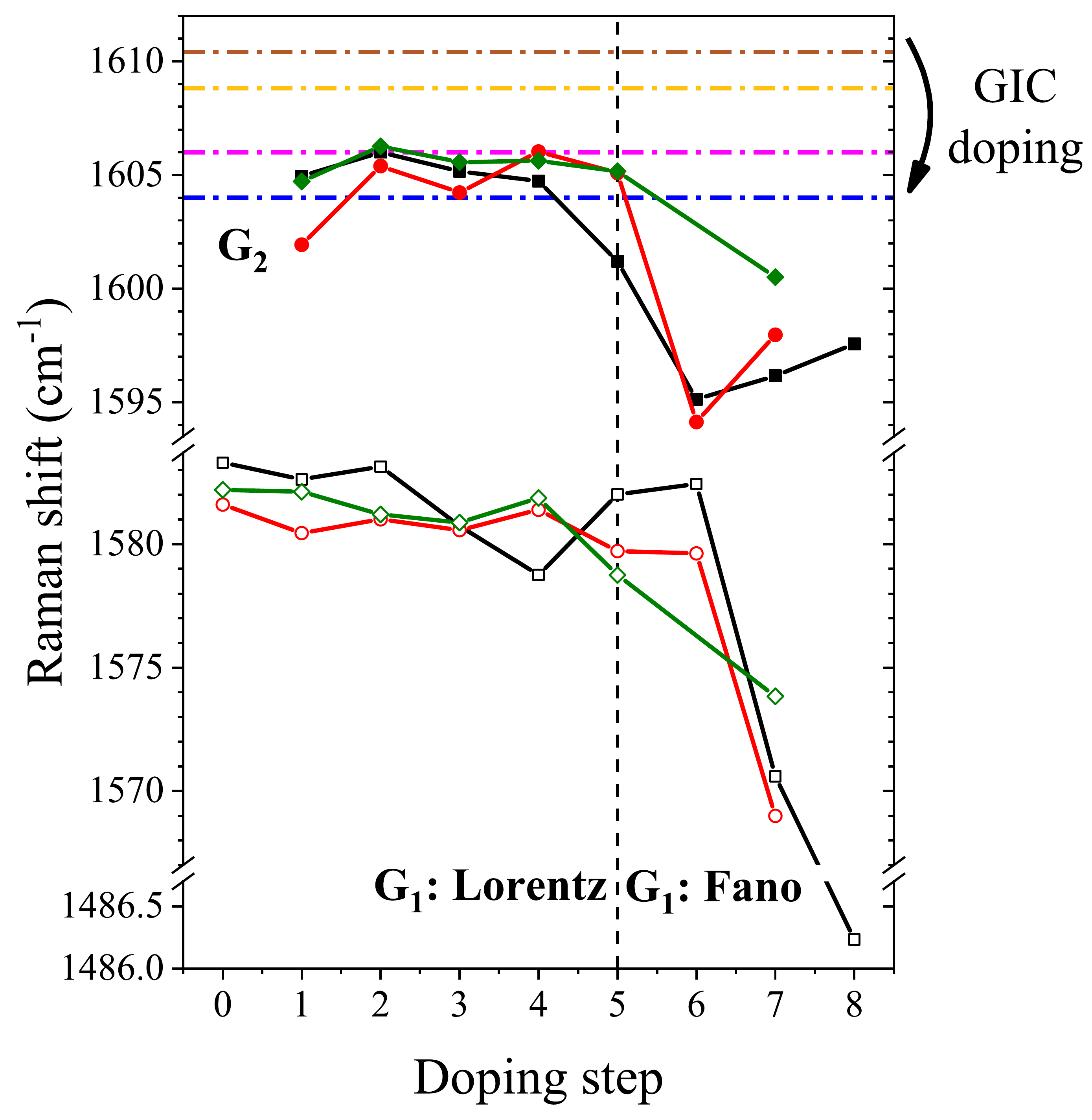}
\caption{Position of the G$_1$ (open symbols) and G$_2$ (filled symbols) Raman modes as a function of the doping step in the investigated FLG species at $514$ nm laser wavelength. The ultrasound treated material is shown with black, the shear mixed one is represented with red and the mechanically stirred sample with green color. The $0$th doping step corresponds to \textcolor{black}{the starting materials}. Positions are obtained through fitting the peaks with Lorentzian and Breit-Wigner-Fano functions, transition between the two shapes is denoted with a vertical dashed line. Relevant G$_{\textrm{c}}$ modes of the potassium intercalated GICs are shown with dashed-dotted lines: KC$_{24}$ (blue), KC$_{36}$ (magenta), KC$_{48}$ (yellow), KC$_{60}$ (mahagony) \cite{ChaconTorres2013}.}
\label{fig:Fig3_g_bands}
\end{figure}

To gain deeper insight into the composition of the multiple restacked FLG, we analyze the G- and 2D Raman-bands. Intercalation step dependence of the split G-bands (G$_1$ around 1580 cm$^{-1}$ and G$_2$ around 1600~cm$^{-1}$) are shown in Fig.~\ref{fig:Fig3_g_bands} for all three investigated types of samples. To understand the origin of each G band, we recall the Raman response properties of potassium doped GICs. Therein, the upper and lower G lines were attributed to charged (the $\textrm{G}_{\textrm{c}}$ band) and uncharged layers (the $\textrm{G}_{\textrm{uc}}$ band), respectively~\cite{ChaconTorres2013}. Upon doping, the $\textrm{G}_{\textrm{c}}$ band moves to lower Raman shift beyond experimental error (horizontal lines in Fig.~\ref{fig:Fig3_g_bands}). The charges transferred from potassium accumulate on the layer immediately adjacent to the potassium layers, which give rise to the $\textrm{G}_{\textrm{c}}$ band. Charge transfer to the rest of the layers (the so-called inner layers) remains low and varies with stage numbers of the GIC. We note that the $\textrm{G}_{\textrm{uc}}$ band also shifts slightly between the different stages due to strain effect. 

In FLG, the G$_2$ band arises from charged graphene layers. However, the comparison with the position of the different GIC stages (see Fig.~\ref{fig:Fig3_g_bands}) unveils a markedly different behavior for the G$_2$ band in FLG and the $\textrm{G}_{\textrm{c}}$ band in GIC. Namely, the position of the G$_2$ band is i) independent of the doping steps, ii) its Raman shift position lies between the position of charged G-band in KC$_{24}$ and in KC$_{36}$. This is a strong indication that the G$_2$ band corresponds to graphene layers that appear to be doped as in Stage 2 or 3 graphite. It also means that in our FLG samples, no higher stages (or lower doping levels) can be achieved. Given the heterogeneous nature of the number of layers in the FLG sample, this reveals that our sample is free from flakes with more than 3-5 restacked graphene layers. \textcolor{black}{This observation is in full agreement with our AFM statistical analysis.} 

Fig.~\ref{fig:Fig3_g_bands} shows that the position of the G$_1$-line barely changes as a function of doping as long as a Lorentzian line fits best the Raman line. This exposes that the induced strain is not affected by the doping level, hence, the G$_1$ line corresponds to significant amount of undoped flakes. Presence of these undoped flakes along with appearance of Stage 3 doping in five-layer-thick flakes highlights the important contribution of undoped flakes with smaller thickness (mono-, bi-, tri-, and four-layer ones).

To further emphasize the differences between the FLG and the graphite powder, we extract a measure of the charge transfer, the electron-phonon coupling parameter (EPC). The electron-phonon scattering linewidth can be estimated from the positions of the Fano lineshape using the expression
\begin{equation}
\gamma^{\textrm{EPC}}=2\sqrt{(\omega_{\textrm{Fano}} - \omega_{\textrm{A}})(\omega_{\textrm{NA}}-\omega_{\textrm{Fano}})}.
\end{equation}
Here, $\omega_{\textrm{Fano}}$ is the measured position of the G-line peak, $\omega_{\textrm{A}}$ and $\omega_{\textrm{NA}}$ are the calculated adiabatic and non-adiabatic phonon frequencies \cite{ChaconTorres2012, Saitta2008}. We approximate the latter two quantities with the ones calculated for KC$_8$: $\omega_{\textrm{A}}=1223$ cm$^{-1}$ and $\omega_{\textrm{NA}}=1534$ cm$^{-1}$, as no exact calculation exists for FLG. This approximation was found to be valid in similar hexagonal carbon systems such as potassium doped multiwalled carbon nanotubes \cite{ChaconTorres2016}.
\begin{table}[h!] 
\begin{center}
 \begin{tabular*}{\linewidth}{c @{\extracolsep{\fill}} cccc} \hline \\
  Sample       & $\omega_{\textrm{Fano}}$ & $\Gamma_{\textrm{Fano}}$ & $q$  & $\gamma^{\textrm{EPC}}$ \\ \\ \hline
  FLG step 10  & $1505$                 & $148$                              &$-1.5$& \textcolor{black}{$181$} \\
  SGN18 Stage I& $1515$                 & $89$                               &$-0.7$& $148$ \\
  HOPG Stage I (Ref.~\cite{ChaconTorres2012}) & $1510$                             & $118$                              &$-1.9$& $166$ \\ \hline
 \end{tabular*}
 \caption{Electron-phonon coupling parameters from the analysis of the G-modes. The values of $\omega_{\textrm{Fano}}$, $\Gamma_{\textrm{Fano}}$, and $\gamma^{\textrm{EPC}}$ are in cm$^{-1}$. Calculated parameters in maximally intercalated FLG are compared to values found in graphite powder (SGN18 Stage I), and Stage I HOPG.}
 \label{tab:epc_params}
\end{center}
\end{table}

The extracted values are summarized in Table~\ref{tab:epc_params}. Therein, $\Gamma_{\textrm{Fano}}$ is the linewidth of the Fano lineshape. In accordance with previous findings in GICs~\cite{Saitta2008}, the $\gamma^{\textrm{EPC}}$ of SGN18 and FLG follow the linewidth of the Fano lineshape linearly ($\Gamma_{\textrm{Fano}}\approx\gamma^{\textrm{EPC}}$). Comparison of the measured characteristics reveals that charge transfer is the largest in HOPG and in FLG, followed by SGN18. Weaker charge transfer in SGN18 can be explained by its morphology, as powders are more difficult to intercalate~\cite{Fabian2012}. Thus, the larger charge transfer in FLG in powder form is a remarkable proof of a system with weak internal strain due to majority of one- to three-layer flakes.

\begin{figure}[h!]%
\includegraphics*[width=\linewidth]{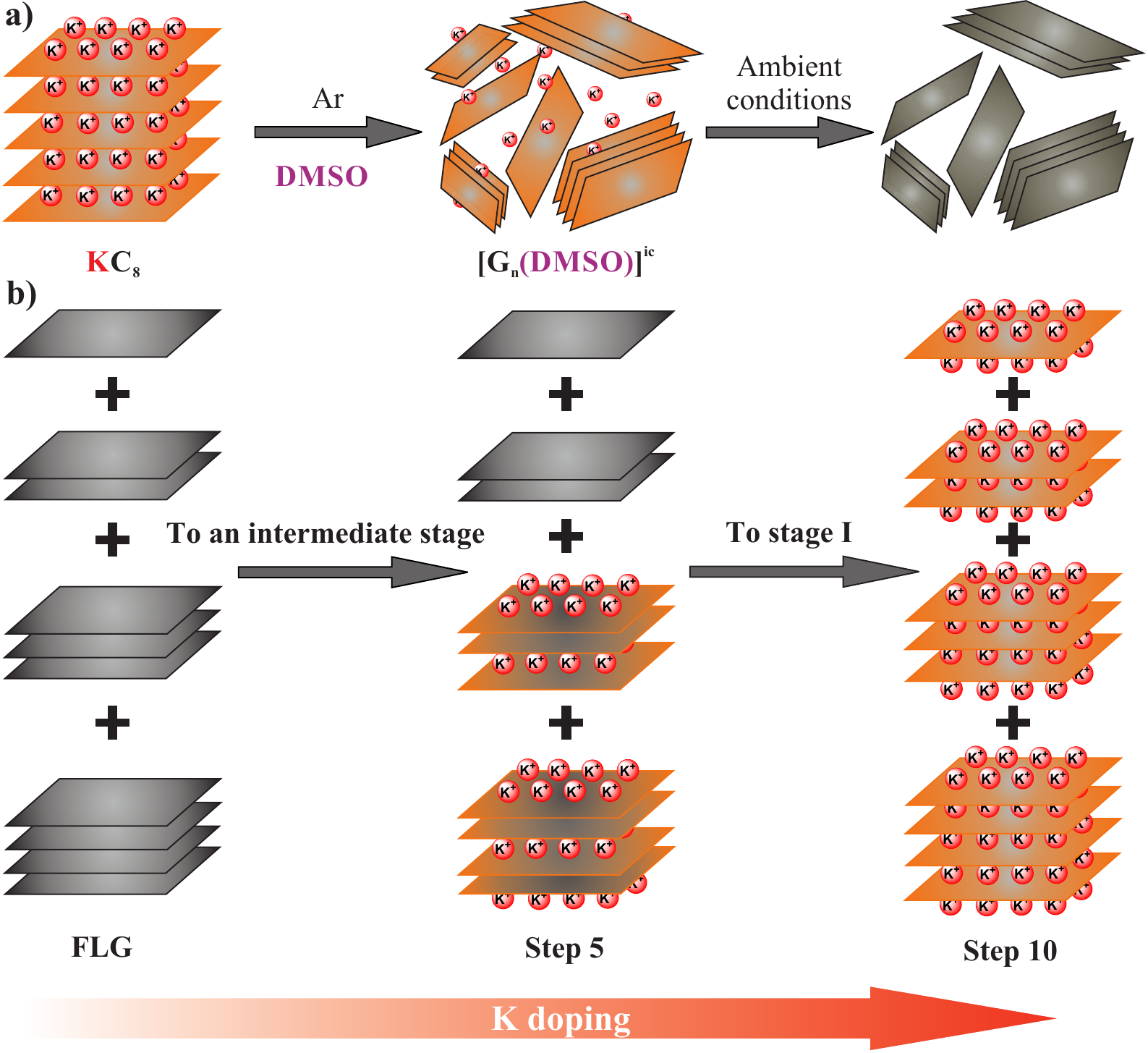}
\caption{Proposed scheme of alkali doping for the FLG sample. a) Synthesis steps of the starting FLG material. b) Illustration of the \textit{in-situ} intercalation process. The sample is a mixture of a few layers: moderate doping affects the flakes with more layers (Steps 1-7) and higher doping steps (Steps 8-10) results in a full doping of all flakes including those consisting of entirely single graphene layers.}
\label{fig:scheme} 
\end{figure}

Figure~\ref{fig:scheme} summarizes the proposed doping scheme for the FLG sample, which allows to gain insight into the heterogeneous layer number distribution. At the beginning of the K intercalation (Steps 1-7), only a high Stage (Stage 3, as we identified) can be reached, which is geometrically possible only in flakes containing restacked graphene of at least $5$ layers. Thus, flakes consisting of less than $5$ restacked graphene layers remain intact from potassium doping at these steps. As the doping proceeds, it is only a low amount of flakes that become intercalated, as strictly speaking our doping steps do not form a material in thermodynamical equilibrium due to the inhomogeneous composition. At higher doping (Step 8-10), flakes with smaller thicknesses start to be doped and eventually all graphene layers are doped to saturation, which corresponds to the structure of Stage 1 GIC.

This scenario is supported by first-principles studies~\cite{Zhu2004,Nobuhara2013}, i.e., that potassium doping yields a formation energy gain ($\Delta F$) that decreases for lower stages up to Stage 2, and increases for Stage 1. This staging phenomenon means that all flakes of the sample reach the same stage before a new stage is started to be formed. A same effect was found experimentally in bilayer graphene individual flakes, i.e., that the doping occurs first on one of the layers reaching a full Stage 2 doping (top layer, in general) before accumulating in-between all layers~\cite{Bruna2010,Parret2013}.

In conclusion, we presented a Raman spectroscopy-based technique to identify the maximal flake thickness in few-layer bulk graphene samples. The presented method is based on studying \textit{in-situ} K doping of FLG samples. \textcolor{black}{Our method uses the combination of the G-band position, its intensity, and the position of the 2D mode components to determine the typically thickest flakes in the sample and to confirm the presence of single-layer ones.} The technique works well on FLG powder samples prepared using wet chemical exfoliation technique and was tested for three different mechanical processing routes. \textcolor{black}{Statistical AFM shows that such samples consist of flakes with a non-uniform distribution of the number of graphene layers, and 90\% of the flakes consist of less than 5~layers}. The Raman spectra of intermediately doped FLG samples can be best described as a sum of two components, corresponding to doped and undoped graphene flakes. The former was argued to arise from five-layer-thick flakes and the latter from flakes made of fewer numbers of graphene layers. \textcolor{black}{Remarkable agreement of our AFM statistical analysis and our Raman data validates our method.}

\section{Contributions}
F.S.~initiated the research, and P.Sz.~suggested the idea. P.V., K.E and J.M.E.~prepared the few-layer graphene samples. U.M.~and K.E.~performed the AFM analysis. B.G.M., P.Sz.~and J.C.C.-T.~carried out the Raman measurements. P.Sz., B.G.M.,~and F.S.~interpreted the data, and laid out the paper structure. All authors contributed to the writing of the manuscript. P.Sz.~and B.G.M.~contributed equally to this work.

\section{Acknowledgment}
Support by the National Research, Development and Innovation Office of Hungary (NKFIH) Grant Nrs. K119442 and 2017-1.2.1-NKP-2017-00001 are 
acknowledged. P.Sz., B.~N., and L.~F.~thank the Swiss National Science Foundation (Grant No.~200021\underline{ }144419). J.~C.~C.-T.~acknowledge the financial support of the DRS Postdoc Fellowship Point-2014 of the NanoScale Focus Area at Freie Universit\"{a}t Berlin. P.~E., K.~E., J.~M.~E., U.~M., A.~H., and F.~H.~thank the Deutsche Forschungsgemeinschaft (DFG-SFB 953 ,,Synthetic Carbon Allotropes'' Project A1) for financial support.


\clearpage
\newpage
\section*{SUPPLEMENTARY MATERIAL}
\setcounter{section}{1}
\setcounter{figure}{0}
\makeatletter 
\renewcommand{\thefigure}{S\@arabic\c@figure} 


\subsection{Color change upon K doping}

\begin{figure}[tb]%
\includegraphics*[width=0.5\linewidth]{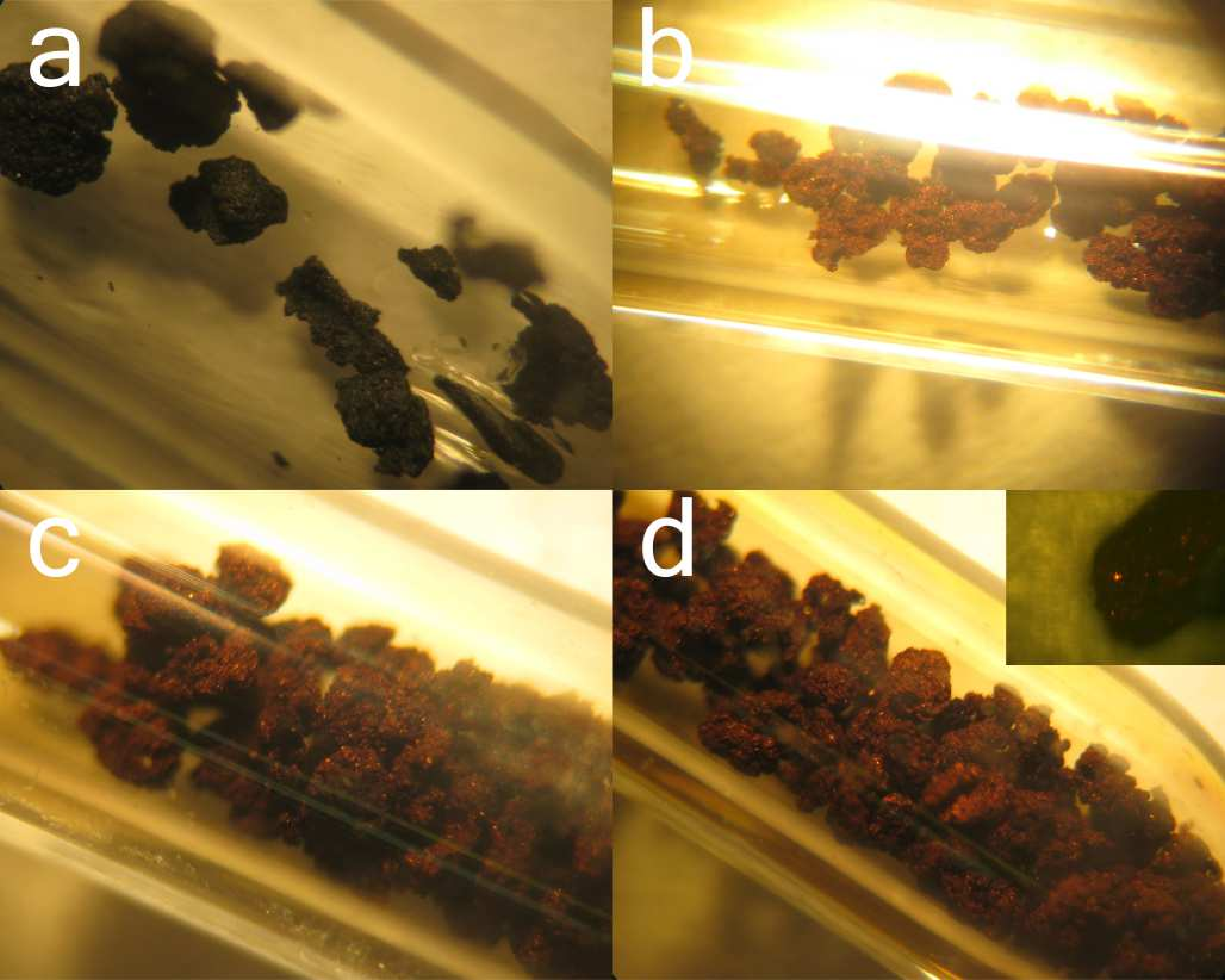}
\caption{Different doping stages: \textit{a)} undoped material, \textit{b)} after 1 hour doping, \textit{c)} after 1.5 hour doping, \textit{c)} after 2 hour doping, \textit{c)} saturation doped material after 3 hour doping. Inset: Microscopic image of an \textit{in-situ} optimally doped sample.}
\label{microscope_images}
\end{figure}

\begin{figure}[tb]%
\includegraphics*[width=0.5\linewidth]{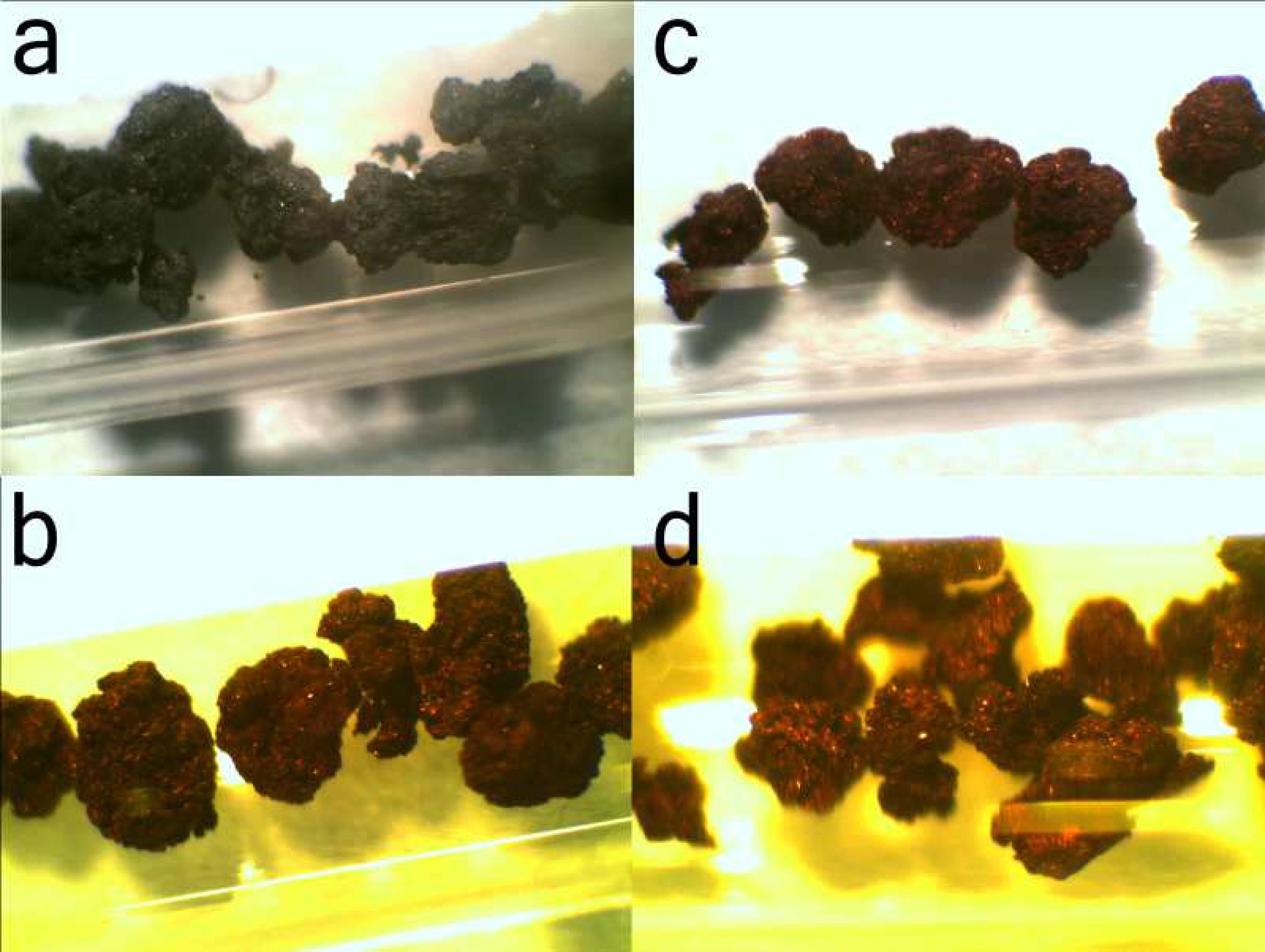}
\caption{Potassium doped FLG in different doping stages: \textit{a)} undoped material, \textit{b)} after 2.5 hour doping at 200 $^{\circ}$C, \textit{c)} after 14.5 hour doping at 200 $^{\circ}$C, \textit{c)} after 19 hour doping at 200 $^{\circ}$C, \textit{c)} after 37 hour doping at 200 $^{\circ}$C and 1.5 hour doping at 300 $^{\circ}$C}
\label{microscope_images2}
\end{figure}

Figure~\ref{microscope_images} displays the color of few-layer graphene upon potassium doping. We identified a gradual color change from gray to deep red as shown in Figure~\ref{microscope_images} (see Figure~\ref{microscope_images2} for another sample). Whereas graphite intercalation compounds show a similar color change (e.g.\ Stage-1 samples with KC$_8$ stoichiometry are gold), no such significant change is apparent for graphite powder samples due to surface roughness \cite{Fabian2012}. This difference is seen as a proof of the smooth surface of the graphene layers in FLG.


\subsection{Analysis of saturation doped FLG}
\begin{figure}[tb]%
\includegraphics*[width=0.5\linewidth]{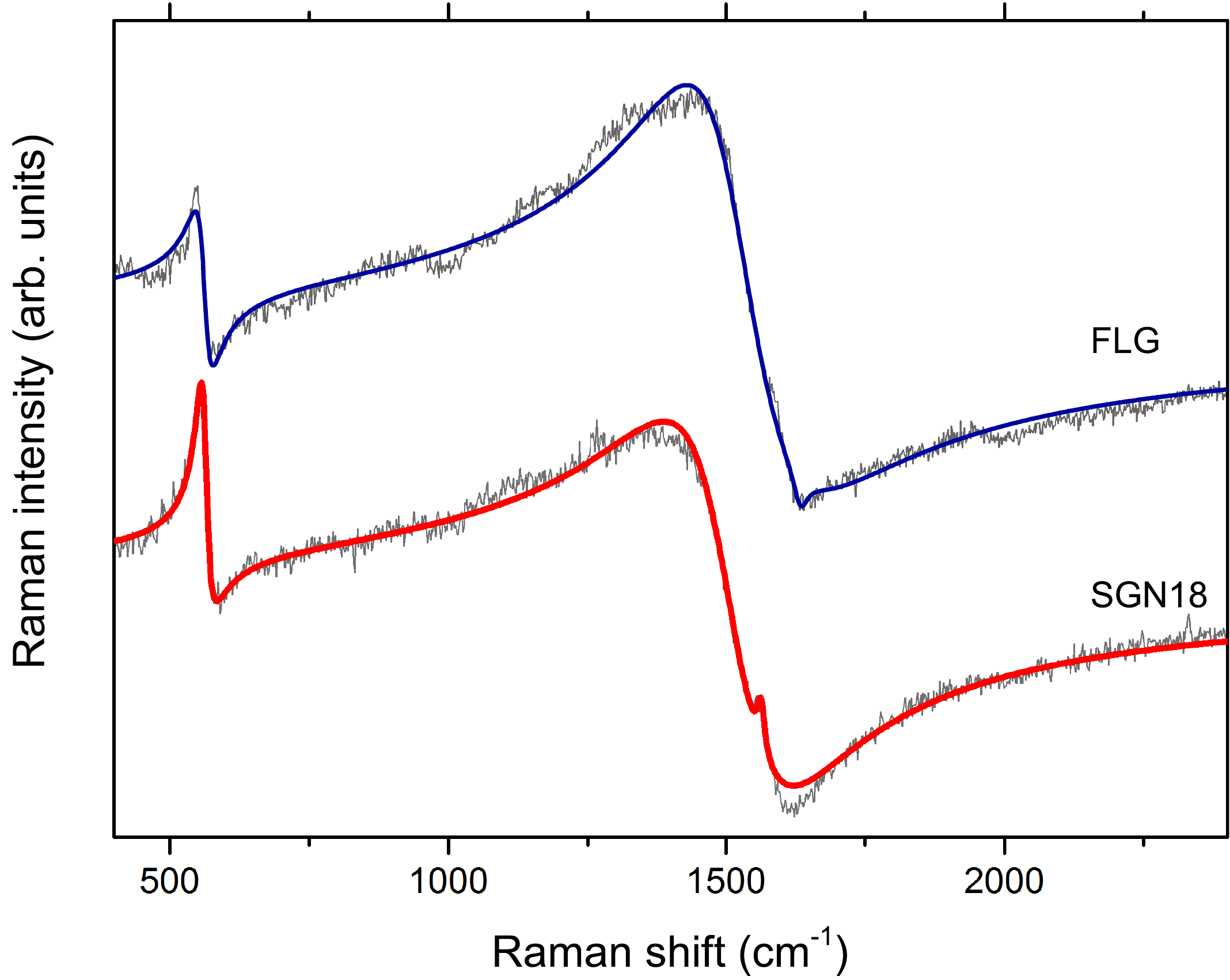}
\caption{Raman spectra of saturation potassium doped ultrasounded FLG and graphite powder. The fit of the analysis is shown as a thick line.}
\label{saturation_doped}
\end{figure}
In Figure~\ref{saturation_doped}, an analysis of the saturation doped reference graphite powder (SGN18) and of the chemically exfoliated sample (FLG) is displayed. Both spectra can be well fitted with the sum of an E$_{\textrm{2g2}}$ Breit-Wigner-Fano component, a Lorentzian component around 1560~cm$^{-1}$, and the C$_{\textrm{z}}$ mode. As discussed in the main text, the appearance of the Fano line at saturation doping is a signature of the presence of non single-layer species in the powder. The Fano components fitted to the two spectra are similar with a phonon frequency of $\omega_{\textrm{ph}}\approx 1515$~cm$^{-1}$, an asymmetry of $q\approx -0.7$, and a damping of $\Gamma\approx 89$~cm$^{-1}$. Within the error bar of our analysis, these values do not differ from those found in KC$_8$. The weak graphitic component around 1560~cm$^{-1}$ is a common feature with the GICs~\cite{ChaconTorres2012}, and it is assigned to the Stage 2 (KC$_{24}$) level doping in FLG and in SGN18.\\
Regarding the \textit{c}-axis mode, the so-called C$_{\textrm{z}}$ mode around 560~cm$^{-1}$, it is present in monolayer, few-layer graphene and in KC$_8$~\cite{Howard2011,ChaconTorres2012}. This M-point mode is only Raman active for high intercalation levels. The large intensity of the C$_{\textrm{z}}$ mode is therefore a manifestation of successful maximal doping in our sample. Whereas the overall intensity in the two samples is comparable, the linewidth and the asymmetry differs in SGN18 and in FLG. The measure of the asymmetry, $\left|1/q\right|$ is larger in FLG, pointing to larger coupling with the electronic continuum. This is plausible as the small number of layer increases the relative doping in FLG. In parallel with this, the broader C$_{\textrm{z}}$ mode can be assigned to the larger vibration energy uncertainty of the system.



\subsection{Statistical analysis of the G-mode positions}
\begin{figure}[tb]%
\includegraphics*[width=0.5\linewidth]{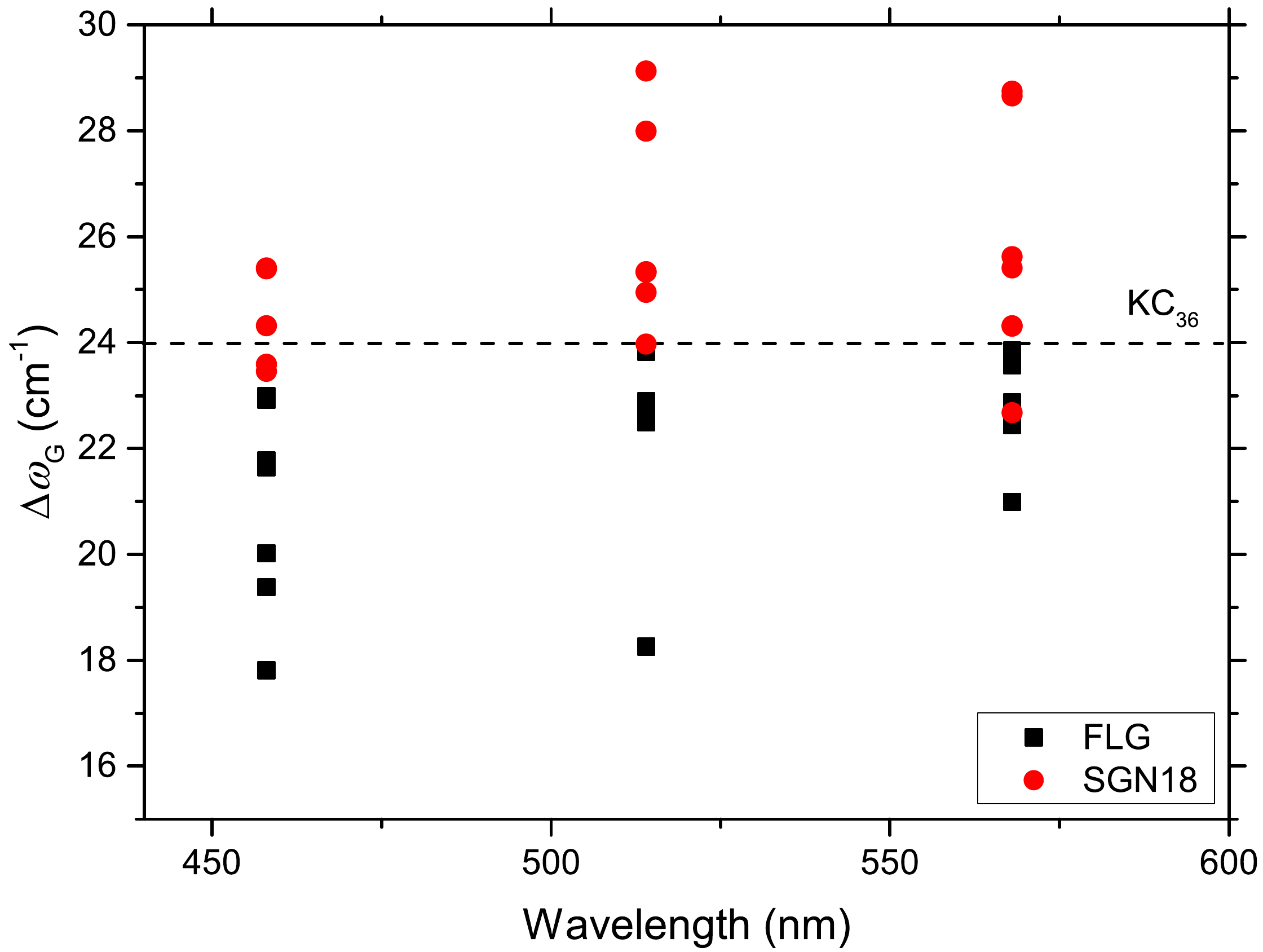}
\caption{Phonon frequency difference of the G$_{\textrm{c}}$ and the G$_{\textrm{s}}$ modes in FLG and in SGN18 at all measured wavelengths and for all doping stages. The dashed line is the Stage 3 single crystal value from Ref.~\cite{ChaconTorres2013}.}
\label{deltag_plot}
\end{figure}

To compare the G modes of the SGN18 graphite powder and that of the chemically exfoliated FLG, we show in Figure~\ref{deltag_plot} the phonon frequency difference of the G$_2$ and the G$_1$ modes ($\Delta\omega_{\textrm{G}}=\omega(\textrm{G}_2)-\omega(\textrm{G}_1)$) for all doping steps and for three different wavelengths in samples prepared with different mechanical processing routes. In order to facilitate the comparison, the $\Delta\omega_{\textrm{G}}$ value in KC$_{36}$ is also shown as a dashed line \cite{ChaconTorres2013}. The Raman shift differences in SGN18 have a minimum at the single crystal value, while it is a maximum for FLG within the error bar of our analysis. Hence, the FLG Raman spectra are, in average, the mixtures of lower G$_2$ and higher G$_1$ modes than those found in SGN18. The lower G$_2$ lines stem from highly charged (close to KC$_{36}$), the higher strained the higher G$_1$ lines arise from unstrained few-layer flakes. This is in perfect agreement with our observation of a two-component Raman spectrum and our decomposition of the Raman signal as shown in the main text.
\subsection{Intensity ratio of G-mode components}

In order to further demonstrate that the Raman spectrum of lightly doped FLG evinces a two-component system, we compare the intensity ratios of the two G lines and the position change of the 2D bands. As we discussed above, the charge transfer happens predominantly from the alkali intercalant to the neighboring graphene layer. This implies that in the GICs, where all the measured sample reaches a constant stage, the intensity ratio of strained and charged G lines ($R$=Int(G$_{\textrm{s}}$)/Int(G$_{\textrm{c}}$)) can be correlated with the stage number. In previous works~\cite{Solin1980,ChaconTorres2013}, a linear increase of $R$ was found as a function of stage number independently of the alkali metal intercalant. Therein, it was shown that it is safe to use the linear dependence of $R$ for the identification of different stages. 

In general, this linear correlation reveals that the number of undoped and doped layers, i.e., the K:C ratio can be determined using the $R$ ratio. Our fit to the G lines of lightly doped FLG Raman spectra indicate an intensity ratio of $R\approx 1$ at $514$~nm, significantly larger than $R\approx0.4$ seen in Stage 6 potassium GIC (KC$_{72}$). Assuming a linear dependence in-between $R$ ratio and the doping level, it evidences an approximate K:C$\approx$1:200 doping level in the first doping step.

In contradiction with the determined doping level, the 2D line exhibits a peculiarly large downshift. As the mode arises from the uncharged inner carbon layers, the 2D line is only present in Stage 3 (KC$_{36}$) and in higher GIC stages. The downshift of the 2D line components can be derived from the gradually increasing biaxial strain due to lattice expansion upon potassium doping \cite{ChaconTorres2013}. The downshift by about 40~cm$^{-1}$ found in KC$_{36}$ is notably smaller than the downshift by about 70~cm$^{-1}$ measured in our lightly doped FLG sample.

Thus, the large 2D downshift is in sharp contrast to the K:C=1:200 doping level found for the G modes. Given that the local strain is expected to be smaller in FLG than in single crystal graphite, this discrepancy can only be explained by the above described coexistence of undoped and doped SLG and FLG flakes.

\subsection{Simulation of FLG Raman spectrum at $\lambda=633$~nm}
\begin{figure}[h!]%
\includegraphics*[width=0.5\linewidth]{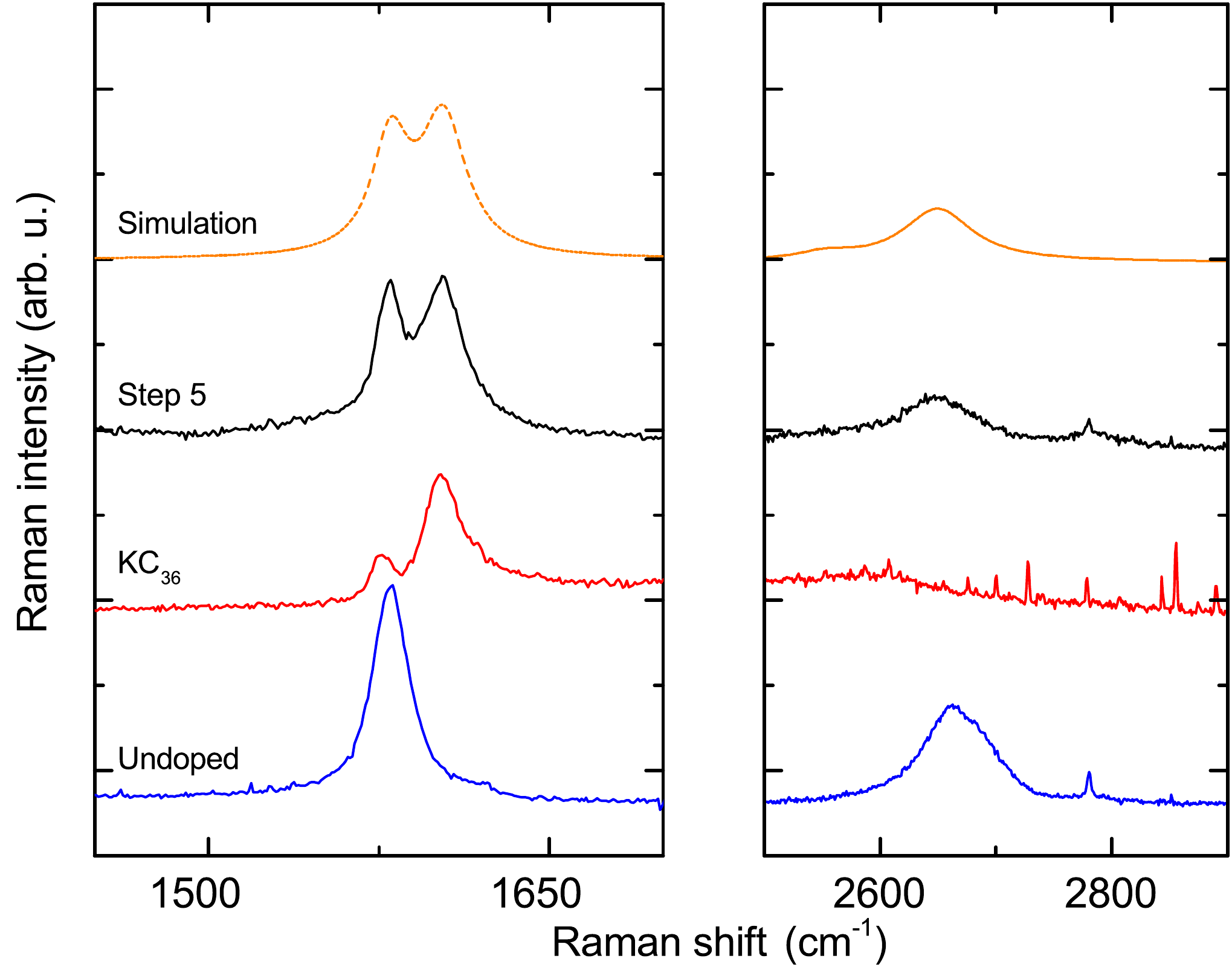}
\caption{Simulation of intermediately doped FLG Raman spectrum as a superposition of undoped FLG and single crystal Stage 3 Raman spectra at $\lambda=633$~nm (FLG Raman spectra) and $\lambda=647$~nm (single crystal Raman spectrum)}
\label{flg_633_647}
\end{figure}

Figure~\ref{flg_633_647} shows a similar blow-up at $\lambda=633$~nm as shown at $514$~nm in the main text. (Single crystal Raman spectrum is measured at $\lambda=647$~nm.) Similarly to the other case, the simulated spectrum (sum of two components) describes well the experimental one. In the Raman spectrum of intermediately doped FLG, the intensity ratio is Int(G$_{uc}$)/Int(G$_{c}$)$\approx0.8$, similar to the ratio at $\lambda=514$~nm, whereas it changes significantly in single crystals~\cite{ChaconTorres2013}.

\subsection{Fitting procedure}
\label{fit_proc}
Individual, symmetric G- and 2D-mode Raman lines were fitted to Lorentzian absorption curves, which were found to be a convenient and reliable method in the case of GICs~\cite{ChaconTorres2013}. Comparison with a more general Voigtian function did not provide any change of the Raman line characteristics.

\end{document}